\documentclass[preprint,5p,times]{elsarticle}
\usepackage[hyphens]{url}
\usepackage{hyperref}
\usepackage{stfloats}
\usepackage{subcaption}
\usepackage{siunitx}

\usepackage{listings}
\usepackage{xcolor}
\lstdefinelanguage[RISC-V]{Assembler}
{
  alsoletter={.}, 
  alsodigit={0x}, 
  morekeywords=[1]{ 
	lb, lh, lw, lbu, lhu,
	sb, sh, sw,
	sll, slli, srl, srli, sra, srai,
	add, addi, sub, lui, auipc,
	xor, xori, or, ori, and, andi,
	slt, slti, sltu, sltiu,
	beq, bne, blt, bge, bltu, bgeu,
	j, jr, jal, jalr, ret,
	scall, break, nop, csrr, csrw
  },
  morekeywords=[2]{ 
	.align, .ascii, .asciiz, .byte, .data, .double, .extern,
	.float, .globl, .half, .kdata, .ktext, .set, .space, .text, .word
  },
  morekeywords=[3]{ 
	zero, ra, sp, gp, tp, s0, fp,
	t0, t1, t2, t3, t4, t5, t6,
	s1, s2, s3, s4, s5, s6, s7, s8, s9, s10, s11,
	a0, a1, a2, a3, a4, a5, a6, a7,
	ft0, ft1, ft2, ft3, ft4, ft5, ft6, ft7,
	fs0, fs1, fs2, fs3, fs4, fs5, fs6, fs7, fs8, fs9, fs10, fs11,
	fa0, fa1, fa2, fa3, fa4, fa5, fa6, fa7
  },
  morecomment=[l]{;},   
  morecomment=[l]{\#},  
  morestring=[b]",  	
  morestring=[b]'   	
}
\definecolor{mauve}{rgb}{0.58,0,0.82}

\usepackage{amssymb}

\journal{Microprocessors and Microsystems}

\begin{document}

\begin{frontmatter}



\title{Design and implementation of a synchronous Hardware Performance Monitor for a RISC-V space-oriented processor}


\author[inst1]{Miguel Jiménez Arribas}
\author[inst1]{Agustín Martínez Hellín}
\author[inst1]{Manuel Prieto Mateo}
\author[inst1]{Iván Gamino del Río}
\author[inst1]{Andrea Fernández Gallego}
\author[inst1]{Óscar Rodríguez Polo}
\author[inst1]{Antonio da Silva}
\author[inst1]{Pablo Parra}
\author[inst1]{Sebastián Sánchez}


\affiliation[inst1]{
        	organization={Space Research Group, Department of Automatics, University of Alcalá},
        	city={Alcalá de Henares},
        	postcode={28805},
        	state={Madrid},
        	country={Spain}}

\begin{abstract}
The ability to collect statistics about the execution of a program within a CPU is of the utmost importance across all fields of computing since it allows characterizing the timing performance of a program. This capability is even more relevant in safety-critical software systems, where it is mandatory to analyze the software timing requirements to ensure the correct operation of the programs. Moreover, in order to properly evaluate and verify the extra-functional properties of these systems, besides timing performance, there are many other statistics available on a CPU, such as those associated with its resource utilization. In this paper, we showcase a Performance Measurement Unit (PMU), also known as a Hardware Performance Monitor (HPM), integrated into a RISC-V On-Board Computer (OBC) designed for space applications by our research group. The monitoring technique features a novel approach whereby the events triggered are not counted immediately but instead are propagated through the pipeline so that their annotation is synchronized with the executed instruction. Additionally, we also demonstrate the use of this PMU in a process to characterize the execution model of the processor. Finally, as an example of the statistics provided by the PMU, the results obtained running the CoreMark and Dhrystone benchmarks on the RISC-V OBC are shown.
\end{abstract}




\end{frontmatter}


\section{Introduction}
\label{sec:introduction}
RISC-V is a CPU architecture which has been receiving a lot of attention in recent years thanks to its advantageous characteristics, namely being an open standard, its modular design, a prominent academic and open source community, and having the freedom to adapt and use it in vastly different fields and applications, from embedded systems to large-scale computing \cite{redmond_risc-v_2021}. This has also sparked the interest of the space industry, both from NASA in the US with the HPSC from JPL \cite{jpl_high-performance_2020}, and in Europe, where there have been some works towards the adoption of RISC-V as the new standard architecture for space \cite{di_mascio_leveraging_2019,wessman_-risc_2021,wessman_-risc_2022}.

In the last three decades, the Space Research Group (SRG) of the University of Alcalá has participated in the development of flight software and hardware for different space missions \cite{meziat_cdpu_1993, sanchez_control_1998, polo_component_2012, sanchez_hwsw_2013}. This experience has motivated different research works oriented to facilitate the fulfillment of the reliability requirements demanded by space missions \cite{da_silva_injecting_2014, r_polo_reliability-oriented_2021, sanchez_arinc653_2021, losa_memory_2023}.
With the same approach, this is why of the whole RISC-V architecture we have focused specially on what allows an implementation to assess its correct behavior and performance, namely for now, its tracing mechanisms \cite{gamino_del_rio_risc-v_2020} and the performance counters.

This article focuses on the latter mechanism. It proposes a solution that facilitates behavior characterization and performance measurement of software deployed on RISC-V based on-board computers (OBCs). Specifically, this work shows a performance measurement (or monitoring) unit (PMU) or, in RISC-V nomenclature, a hardware performance monitor (HPM). Other terms, such as statistical unit or performance counters, are also often used in the literature for this type of unit, so from now on, any of these names will be used interchangeably.

This PMU has been integrated into a RISC-V soft-core on-board processor for FPGA with a segmented pipeline targeting space applications. The PMU supports the events standardized by the RISC-V specification \cite{waterman_risc-v_2019} and allows its extension with additional events both during synthesis and in execution time. It presents a very extensible design, capable of accepting new events with a minimal development cost. Additionally, it offers the advantage of extracting timing information from the measured events, since the counting process is synchronous with the execution of the instructions. This improves the value of the PMU, as it is now possible to see when each event occurs and how they relate to each other.

The primary use case that benefits from the implementation of a PMU is, during the development process, to facilitate the design, development, and debugging of the software, and even of the hardware itself; since with it we can collect the properties of the architecture and correct or optimize its behavior. Typically, after launch, having passed the commissioning state, the HPM can be disabled, to reduce the footprint of the on-board computer (OBC). However, as we shall see throughout the rest of this article, it could be argued that in this case, the increase in resource and power utilization with the PMU enabled is sufficiently small that, depending on the circumstances, it could even be considered advantageous to keep it activated in order to debug problems during flight.

The remainder of this paper is structured as follows: first, in Section \ref{sec:related_works}, a brief overview on the state of the art is outlined. Then, in Section \ref{sec:development}, the main characteristics of the PMU design and its implementation details are discussed, along with the motivation and rationale of their selection. Here, a brief example of operation is also described. Subsequently, in Section \ref{sec:model}, using the data obtained with the PMU, as evidence of its usefulness, the execution timing behavior of our processor is characterized. Next, in Section \ref{sec:results}, the results obtained with the HPM during the execution of different benchmarks used to test the correct operation of the proposed design are presented. Additionally, the results are compared with those of another processor, to externally validate the design and to illustrate the enhancements with other PMU implementations. Moreover, the resource utilization and power measurements of this implementation, in comparison with the same developed OBC but without the hardware performance monitor enabled, are also shown. Finally, in Section \ref{sec:conclusions}, the conclusions are drawn.

\section{Related Works}
\label{sec:related_works}

As stated in the introduction, there has been a significant increase in interest in utilizing the RISC-V architecture for space applications in Europe over the last 5 years \cite{di_mascio_leveraging_2019}. This attention has now escalated beyond purely academic interest due to its advantageous features and now includes economic incentives. For instance, Cobham Gaisler has developed an alternative to the standard OBC in the European space industry, LEON, with a new processor based on RISC-V called NOEL-V \cite{gaisler_product_2022}. Furthermore, the European Union has created various projects within its H2020 framework for the development of the necessary infrastructure for the maturation of this architecture in the European safety-critical and space landscape \cite{eic_-risc_2022,epi_press_2021}.

In addition to its favorable economic outlook, the appeal of RISC-V is also based on its technical characteristics. Specifically, in the area that concerns us, RISC-V allows quite a scalable design for the PMU \cite{waterman_risc-v_2019,waterman_risc-v_2021}. It includes 32 counters out of which only 3 have a standardized purpose, with the other 29 capable of being configured to measure any event needed, both at runtime and during hardware synthesis. These 29 counters can be configured via control registers, each of which allows at least $2^{32}$ selectable events for a single counter, a value that exceeds any practical implementation, producing considerable design flexibility.

Regarding other architectures, current multipurpose processors like Intel's \cite{m_johnson_intel_2010} or ARM’s \cite{arm_cortex-a5_2009} include proprietary PMUs. These units still primarily only measure internal microprocessor events, not system-level tasks, or other software constructs \cite{salapura_next-generation_2008}, even though they are much more complex than current embedded architectures like, for example, LEON's \cite{gaisler_grlib_2023,ho_hardwaresoftware_2014}.

In general, most PMU designs have two main mechanisms, an event detector, and the event counters \cite{sprunt_basics_2002}. With this type of system, it is possible to create performance profiles with which to analyze the performance and behavior of the processor and the program. These are called time-based profiles and event-based profiles \cite{sprunt_basics_2002}. Nevertheless, with this design of PMU it is difficult to obtain timing data on when each event occurred and how these events are related to each other \cite{sprunt_basics_2002,dean_profileme_1997}. This is the reason why the design presented in this paper synchronizes its counting process with the execution of instructions, as will be discussed in more detail in the following section.

Finally, in addition to the previously mentioned specification, some further works have been conducted regarding the HPM of RISC-V. Firstly, it was analyzed regarding its use with the currently available open-source tools. With this, recommendations were made on how the spec and surrounding software should be improved \cite{domingos_supporting_2021}. And secondly, other works have deepened the flexibility of the specification, improving its task awareness and in general the ease of configuration \cite{scheipel_system-aware_2017, lei_highly_2020}.

With all of the aforementioned academic work regarding the HPM of RISC-V, and the industry’s adaptation of the Linux kernel and other open-source software infrastructure to RISC-V, an updated standard \cite{risc-v_risc-v_2021} has been produced. This update provides clarification and some pending functionality which has been long requested, mainly the possibility of launching interrupts when one of the counters suffers an overflow, and the filtering of events depending on the privilege level. Still, even after these updates, we believe that our work is still of interest as no other work that we have found explains in detail the integration and implementation of the PMU within a pipelined architecture. In addition, our design also benefits from the features outlined above, as will be further explained in the next section.

\section{Development}
\label{sec:development}
The implementation presented here was born out of the interest of our research group in finding ways to improve the observability, testability and reliability of processors for space applications.

We have gathered considerable experience in this area, having been in charge of creating the hardware and software for the ICU \cite{r_polo_reliability-oriented_2021,sanchez_hwsw_2013} for the EPD instrument \cite{rodriguez-pacheco_energetic_2020,prieto_-flight_2021} inside the Solar Orbiter mission. In the last 3 years, we have been working on a RISC-V processor where we could instill our experience. Therefore, as mentioned previously in Section \ref{sec:introduction}, we have been focusing on tracing \cite{gamino_del_rio_risc-v_2020} and performance measuring, to be able to gather the behavior of the architecture and ease the development and debugging process of software.

Looking specifically at the PMU, this design was specially created with three goals in mind: firstly, to allow the monitoring of events without interfering with the behavior of the existing pipeline, and thus also transparent to the execution time; secondly, to synchronize the event count with the execution of each instruction so that the maximum observability could be achieved; and thirdly, to support the extensibility of the events that the specification proposes, so that new ones could be added with minimal development cost. 

Hence, in the following subsections the development of this PMU is presented. Initially, the baseline processor is introduced, which served as the starting point for the proposed design. Then, in the following subsection, the PMU design and its motivations are detailed. Subsequently, in the next two subsections, its configuration, and some of its implementation details are shown based on a RISC-V architecture \cite{waterman_risc-v_2019} pipelined OBC. Additionally, a specific example of operation across the entire pipeline is described in the final subsection.

It is important to clarify that while the objective of developing the PMU is to support reliability, its primary function is to provide developers with the essential tools for monitoring and analyzing the processor's behavior. As such, the PMU serves as a foundational component, offering insights into execution patterns and performance metrics vital for diagnosing potential issues and optimizing software performance. However, it is crucial to note that the PMU, in isolation, does not directly deliver reliability. Achieving reliability in processor systems typically necessitates a multifaceted approach, encompassing fault tolerance mechanisms, error correction techniques, and rigorous testing protocols. Thus, while the PMU lays the groundwork, contributing significantly to the development process by enhancing observability and testability, its immediate impact on reliability is indirect, being part of a broader strategy aimed at cultivating reliable and robust processor systems for space applications. Nonetheless, it could be argued that, in comparison with other PMUs, the proposed design offers superior reliability, a topic extensively examined in Subsection \ref{sec:design}.

Lastly, even though for this paper the focus has been on the RISC-V architecture, since the basis of the PMU presented here has been what is defined in its specification, it should be noted that the actual design principles really is ISA-agnostic and could therefore be applied to other CPU architectures interchangeably, as will be demonstrated in the next subsections.

\subsection{Baseline processor}
\label{sec:baseline}

\begin{figure}[ht]
	\centering
	\includegraphics[width=0.85\linewidth]{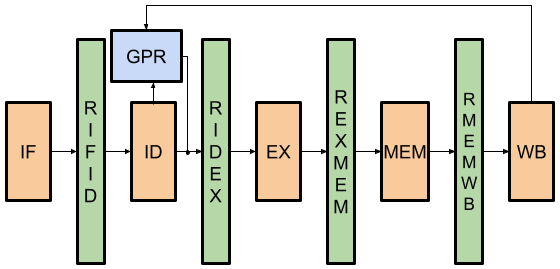}
	\caption{Baseline pipeline structure. The 5 stages mentioned are shown in orange, while in green, the inter-stage registers that synchronize the data transition between them are found. Finally, in blue, other functional units are shown. Specifically, the placement of the general-purpose register file can be seen.}
	\label{fig:baseline-pipeline}
\end{figure}

Before focusing on the design of the PMU a brief description of the baseline pipeline from where the design evolved will be explained. In Figure \ref{fig:baseline-pipeline} a schematic of this baseline architecture can be seen. In it, a typical 5-stage pipeline design can be observed, between which the inter-stage registers manage the propagation of the results, enabling the synchronization of the pipeline. In addition, although for simplicity and ease of understanding they are not represented in the figure, there are some additional components in charge of other functions of the pipeline, e.g., a data forwarding unit (FU) for overcoming data hazards and some additional units for memory accesses and exception flow i.e., traps, etc.

Briefly, the functionality of each stage is the following: the Instruction Fetch stage (IF) controls the insertion of instructions into the pipeline from the instruction memory; the Instruction Decodification stage (ID) recognizes each instruction, detects data hazards, and supplies the appropriate control signals to the remaining stages; the Execution stage (EX) calculates the results needed for the following stages, for instance, it is also here where, in case of branch instructions, it decides whether or not to take them and resolves the jump address. Next, the Memory stage (MEM) is in charge of storing and loading data to and from data memory. And finally, the Write-Back stage (WB): oversees which data, if at all, is stored back in the GPRs, be it the result from the EX stage or the value gathered from memory in the MEM stage.

\begin{figure*}[hbp]
	\centering
	\includegraphics[width=0.93\linewidth]{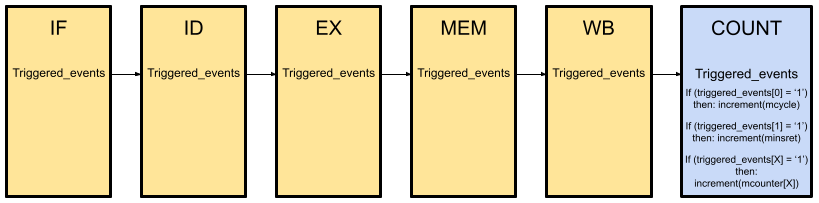}
	\caption{This is a simplified abstraction that shows the coupling between each of the five stages of the pipeline and their corresponding inter-stage registers, hence the difference in color coding with the amalgamations in yellow. In addition, it also shows the \textit{triggered\_events} data structure and how it is monitored and chained through the pipeline arriving to the count module where the events are finally added up. This module is an auxiliary functional unit and therefore not an actual stage in the pipeline since it does not produce any effect in the execution of instructions, thus the blue color.}
	\label{fig:simplified-pipeline}
\end{figure*}

Gamino et al. in \cite{gamino_del_rio_risc-v_2020} give a more detailed look of the whole processor and its distinctive features. From this point onwards the focus will solely be on the design of the PMU. Nevertheless, it is important to remember the segmented nature of the processor, as this is critical for the comprehension of the design.

\subsection{PMU design: motivating factors and proposed solution
}
\label{sec:design}

The PMU design depicted in this paper comprises two main innovative approaches: its decentralized triggering system and its synchronized counting process. The first approach is provided by the most important data structure of the PMU design i.e., ``\textit{Triggered\_events}'', which stores whether each of the supported events has been detected on any of the units composing the pipeline, and, as shown in Figure \ref{fig:simplified-pipeline} and Figure \ref{fig:full-pmu}, it is chained through the pipeline until it reaches the Control and State Registers (CSRs) where the events are finally counted.

This could be considered a decentralized PMU design, and it has the advantage over centralized designs in that events are triggered at the pipeline stage where they arise, rather than feeding all the control signals to the counter unit and triggering the events in the same unit where the counting occurs. A fully centralized design is rare, as the amount of information needed to be fed would increase complexity exponentially. Nonetheless, hybrid designs are quite common, as can be seen on Gaisler’s GR740 board \cite{gaisler_l3stat_2023, gaisler_leon4_2015, gaisler_ahb_2015}.

\begin{figure}[ht]
    \centering
    \includegraphics[width=1\linewidth]{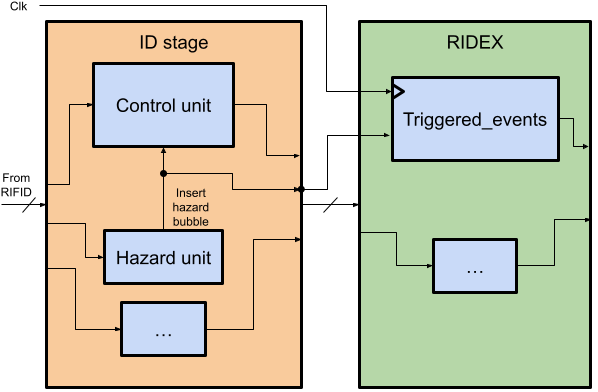}
    \caption{Example illustrating the integration into the existing processor of the hazard event detection mechanism and its storage under the proposed design. As can be seen, the signal “Insert hazard bubble” was already necessary so that the control unit knows when to insert a bubble. Therefore, the PMU mechanism monitors existing signals within the design and registers them in the \textit{triggered\_events} data structure. Notably, this process occurs in parallel without introducing sequential logic, ensuring no timing penalty.}
    \label{fig:event-detector}
\end{figure}

Instead, the design proposed here is entirely decentralized. Therefore, events are triggered across the pipeline, in the same unit where they are detected, and then chained through each remaining pipeline stage, from the Instruction Fetch, i.e., the start of the pipeline, to the Write-Back, i.e., the final stage. This approach simplifies the detection logic, as shown in more detail in Figure \ref{fig:event-detector},  decreasing its complexity, and thus, facilitates the extensibility of the events supported.

The other major contributing aspect of this design is the synchronization of the counting process by modifying the moment at which the count is performed. In other CPUs, in most cases it is impossible to accurately attribute an event to a specific instruction. The difficulty arises because events are counted within the monitoring unit immediately after they have been triggered. Consequently, if events occur at different stages of the pipeline, each event is counted at a distinct phase of its execution, without actually being synchronized with the completion of the instruction execution. As a result, precisely attributing an event to the instruction that generated it becomes challenging. Moreover, there is a risk of potentially counting an erroneous event which would have been canceled in subsequent stages. 

A typical example of this is the event of instruction retirement. There are CPUs which increment this type of event the moment the instruction enters the execution stage, regardless of whether this instruction completes execution or if an error is encountered while still in the pipeline and an exception is thrown \cite{nolting_neorv32_2024}. 

One possibility to solve this problem would be making the event detection logic more complex, as will be discussed in subsequent paragraphs, controlling all the possibilities of event cancellation in the detection logic. But this still would not solve the issue with counting on a different stage as when the instruction is finally committed to the register file, which can still pose significant challenges.

Another even more extreme example arises while utilizing the Sscofpmf extension \cite{risc-v_risc-v_2021}, that allows the generation of interrupts when one of the PMU counters overflows. One could imagine a scenario where different subroutines were incrementing a specific PMU counter. The case may arise where instructions of multiple subroutines were found in the pipeline when an overflow interrupt of that counter is produced. Due to the previously mentioned problem, the latency within the CPU between triggering the event and executing the instructions, the program counter (PC) of the instruction delivered to the interrupt handler may not be the one that caused the event. In fact, the difference between them can vary by an unpredictable amount, as it could be the PC of any of the instructions of any of the subroutines which were incrementing the counter. One could argue, depending on the implementation of the detection logic, that for in-order processors the PC must belong to the instructions within the pipeline, but out-of-order processing exacerbates this problem. This discrepancy could result in faulty information being received by the interrupt handler, potentially leading to undefined behavior. 

Importantly, these kinds of problems are possible with any type of event, depending on the design of the PMU, not only with those in the examples provided, and although they are more common in out-of-order processors, problems can also be found among in-order processors. The works \cite{sprunt_basics_2002} and especially, section 2.2 of \cite{dean_profileme_1997}, discuss more about the intricacies of synchronizing the event counting with the instruction retirement.

This design aims to resolve these problems by instead of storing trigger information within the pipeline, synchronizing event counting with instruction retirement. In Figure \ref{fig:simplified-pipeline}, a simplified abstraction of the pipeline can be seen, where each rectangle represents the fusion between the stage and the inter-stage registers. This figure also shows that the counting of the events is set to only occur on the next cycle after the instruction has finally been written back to the registers, either the General-Purpose Registers (GPRs) or the CSRs.

While, at first glance, it might seem a mistake to account for events later than they happen, this is a side effect of the fact that the events can appear at any moment during the execution, as has been explained. For example, the case may arise that the retirement of an instruction is set to be counted and during the WB stage it is detected that the result cannot be written. Thus, the events triggered by this instruction need to be updated before being recorded on the PMU. Hence, it is necessary to wait until the completion of its execution, i.e., the next cycle after the write-back stage, to finally count all the events.

In general, there are various other ways that the event triggering could be synchronized. For example, Nam Ho et al., in \cite{ho_hardwaresoftware_2014, ho_towards_2014} receive the event pulses and then it is the control logic within the PMU module itself what manages when the counters are finally incremented. Meanwhile, other more complex CPUs, like for example the ones which support Out Of Order and speculative execution \cite{bakhvalov_performance_2020, dean_profileme_1997}, confront this problem via counting the events triggered immediately and waiting until the end of the execution in case any problem was encountered, to either commit the results, or instead, undo the produced events and any other type of side effects generated. Given the complexities and limitations of these latter approaches, they were deemed a hindrance and the simpler design with the considerations discussed in the preceding paragraphs was opted for, as any kind of benefit was outweighed by the drawbacks.

An in-depth examination of the integration of the design within the existing processor is illustrated in Figure \ref{fig:event-detector}. This PMU design involves capturing signals for each event from each of their respective units and, rather than directly routing them to the counter unit as commonly practiced in other PMU designs, concatenating them through the pipeline. Considering this design, several noteworthy aspects emerge. Firstly, there is no timing penalty incurred during instruction execution, as the standard pipeline remains unaltered. The only aspect which is delayed is the arrival of events to the counter unit, but as explained previously this is deliberate, as this way each event can be attributed to the corresponding instruction which is completing execution. Secondly, although the size of the inter-stage registers obviously increases, as will be elaborated in more detail on Section \ref{sec:utilization_power}, this difference is negligible, as it only increases one bit per tracked event per register. Therefore, the expansion in size is minimal in comparison with the value of the information provided. Thirdly, with respect to the critical path potentially impacting processor frequency, as depicted in Figure \ref{fig:event-detector}, the modifications solely add changes in a parallel manner. There is no additional sequential logic that could elongate the path the signals could take with more operations; instead the existing value is simply channeled into a new parallel register. This strategic inclusion of parallel registers for event tracking ensures that the signal processing speed remains unaffected. Consequently, the processor's ability to maintain its performance metrics, despite the augmented functionality of the PMU, remains intact.

Finally, regarding support of this PMU design for more complex architectures, it is pertinent to acknowledge that while the primary objective of this article is to present and validate the core functionality, some consideration has been directed towards potential enhancements for advanced architectures. As mentioned previously, one of the key aspects of the proposed design is that it is architecture agnostic. The performance monitoring hardware is only intrinsically linked with the pipeline, as it is here where it detects and chains the occurred events, rather than with the decodification and execution of instructions, i.e.: the ISA. As such, the development and behavior of this design in the context of more advanced microarchitectures would remain equivalent.

For instance, in pipelined microarchitectures featuring multi-cycle or even out-of-pipe functional units, like for example those common on multiplication or division operations \cite{marena_risc-v_2019}, the instructions would still reach the MEM and WB stages, and therefore its inter-stage register. Then, once these instructions would arrive at this register, the events occurred during its execution would trigger the same way as already described, and when the instructions were committed to the register file their corresponding events would be counted.

Similarly, in architectures employing superscalar or out-of-order execution, although their microarchitectures are more intricate, they still rely on inter-stage registers for signal  synchronization, so the newer events from these architectures could be detected there. And although, in out-of-order processors  execution may occur out of order, the actual commitment of instructions still adheres to architectural order using register renaming. Therefore, consideration should be given to associating events with instructions based on the real physical registers rather than the logical architectural ones during instruction commitment. Furthermore, regarding multi-core architectures, each core or hart, in the usual RISC-V terminology, would possess its own set of events and performance counters, enabling independent monitoring and analysis tailored to each core's activity.

Lastly, the treatment of cache memories has not been considered in this work. It is widely recognized that cache memories introduce inherent uncertainties in the pipeline flow, mainly due to their hit/miss ratio, cache memory configuration and software coding. These uncertainties can make the WCET difficult to estimate  \cite{prieto_leon2_2007, zhang_precise_2022}. Therefore, since the primary objective of this paper is to characterize and validate the core functionality of the presented HPM for RISC-V, the inclusion of cache memories would have introduced additional complications and, as a first approximation, for the time being, they have not been implemented. Consequently, the usual cache hits and misses events are not yet supported, as detailed in the next subsection. Nevertheless, increasing processor performance stands as an imperative in the contemporary landscape,  both for the old and for the new space paradigms. Hence, providing execution acceleration mechanisms (cache, multicore, etc.) support is one of the first improvements planned. For example, implementation strategies could mirror those employed in architectures like Intel's Nehalem, where caches are embedded into the pipeline \cite{thomadakis_architecture_2011}, facilitating that events could be automatically detected and triggered in a process equivalent to what has already been explained. Thereby, requiring no modifications to the PMU design philosophy. In cases where support for caches beyond L1 is warranted, such as L2 or L3 caches, the fetch stage would assume responsibility for managing events generated by these caches via additional communication lines. Then, after an analysis to characterize the timing behavior of the cache, such as the one that can be seen in \cite{prieto_leon2_2007}, these will be integrated into our OBC and the actual events supported by the PMU.

\subsection{PMU Configuration}
\label{sec:configuration}

\begin{figure*}[htb]
	\centering
	\includegraphics[width=0.87\linewidth]{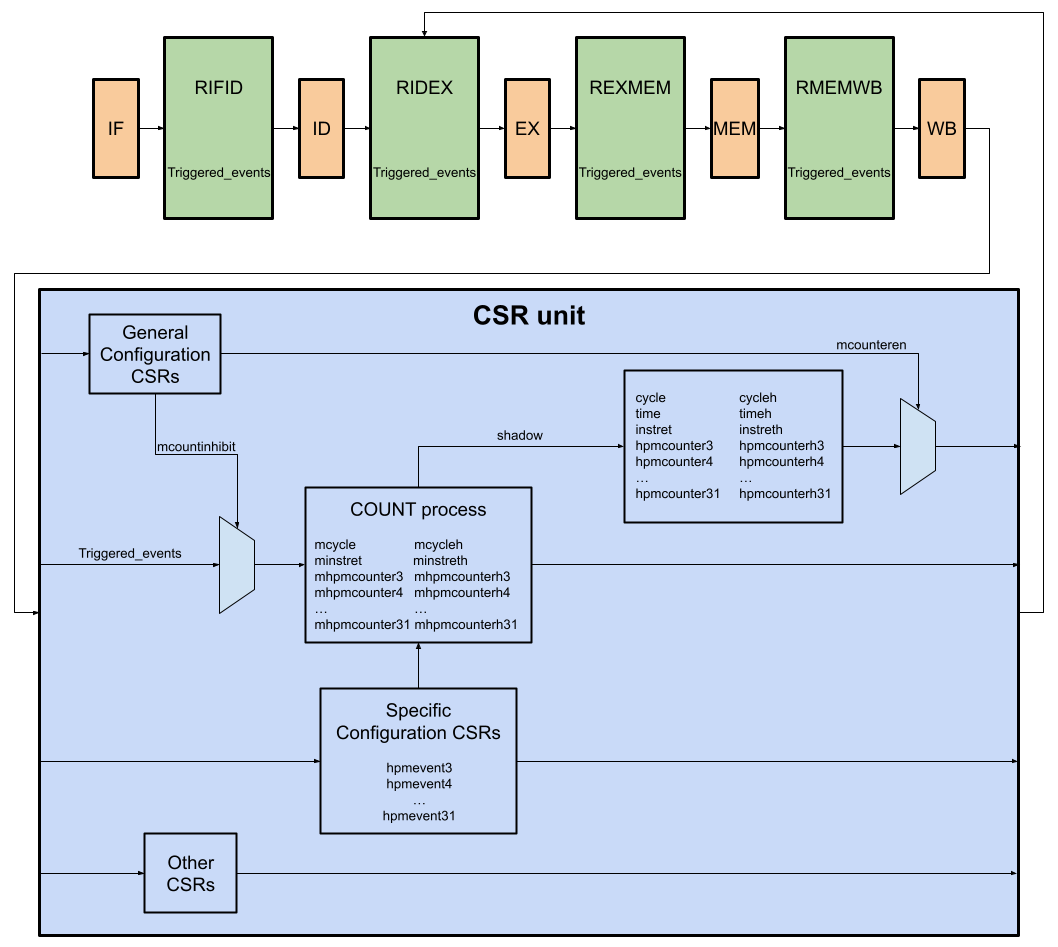}
	\caption{Here the complete diagram of the PMU is shown. As has been explained, its reach spans throughout the entire pipeline by storing the monitored events in the \textit{triggered\_events} data structure, which ultimately arrives at the CSR unit. Then it is here where the behavior of the PMU is decided with its configuration registers, and where the performance counters are located and finally incremented.}
	\label{fig:full-pmu}
\end{figure*}

In Figure \ref{fig:full-pmu}, the full PMU design can be appreciated. The figure shows how the PMU spans across the whole pipeline, with the events triggered in each stage being chained through the remaining stages and finally reaching the CSR unit. Furthermore, it also shows where the configuration registers are located in order to decide the functionality of the PMU and where the counters are incremented.

Once the PMU behavior has been described now these configuration registers will be explained. As mentioned earlier, this configuration could be changed at any moment, both during synthesis and at runtime. According to the scope of their function, these registers could be separated into general and specific configuration registers, and they are the same as those defined in the RISC-V specification \cite{waterman_risc-v_2019}. On the one hand, the general configuration registers are \textit{mcounteren} and \textit{mcountinhibit}. The former decides whether each counter is accessible in both user and machine mode or, instead, they are only accessible in machine mode. Whereas the latter register inhibits the counting of the bit-selected counters. This can be useful in several ways, such as to reduce power consumption, or to allow access to the entire set of counters without their value changing.

Meanwhile, on the other hand, the specific configuration registers (\textit{mhpmevent3 - 31}) oversee the selection of which event is monitored by which counter so that when triggered, an increment is fulfilled in the selected counter. The supported events can be found in Table \ref{tab:events}. This table also depicts the aforementioned relationship between the counters and the event selected for each counter.

\begin{table}[h]
\centering
\resizebox{\columnwidth}{!}{
\begin{tabular}{|c|c|c|}
\hline
Counter   	& \begin{tabular}[c]{@{}c@{}}Configuration\\register\end{tabular} & Programmed event     	\\ \hline
mcycle    	& -                                                           	& HPM\_EVENT\_CYCLE    	\\ \hline
-         	& -                                                           	& -                    	\\ \hline
minstret  	& -                                                           	& HPM\_EVENT\_INSTRET  	\\ \hline
mhpmcounter3  & mhpmevent3                                                  	& HPM\_EVENT\_EXCEPTION	\\ \hline
mhpmcounter4  & mhpmevent4                                                  	& HPM\_EVENT\_EXT\_INT 	\\ \hline
mhpmcounter5  & mhpmevent5                                                  	& HPM\_EVENT\_TIME\_INT	\\ \hline
mhpmcounter6  & mhpmevent6                                                  	& HPM\_EVENT\_BRANCH   	\\ \hline
mhpmcounter7  & mhpmevent7                                                  	& HPM\_EVENT\_BRANCH\_NT   \\ \hline
mhpmcounter8  & mhpmevent8                                                  	& HPM\_EVENT\_UNCOND\_JUMP \\ \hline
mhpmcounter9  & mhpmevent9                                                  	& HPM\_EVENT\_HAZARD   	\\ \hline
mhpmcounter10 & mhpmevent10                                                 	& HPM\_EVENT\_MEM\_ACCESS  \\ \hline
mhpmcounter11 & mhpmevent11                                                 	& HPM\_EVENT\_LOAD     	\\ \hline
mhpmcounter12 & mhpmevent12                                                 	& HPM\_EVENT\_STORE    	\\ \hline
mhpmcounter13 & mhpmevent13                                                 	& HPM\_EVENT\_FETCH    	\\ \hline
\end{tabular}
}
\caption{Table showing the matching between each counter, its configuration register, all the currently supported events and which counter counts each event.}
\label{tab:events}
\end{table}

Now a brief description of each event will be provided. First, the cycle event counts the number of cycles that have elapsed since its reset, and it is triggered and incremented on every rising edge for as long as its corresponding counter is not inhibited. On this note, outside the main unit of the PMU (as it is outside of its definition in the specification \cite{waterman_risc-v_2019}, though it can be accessed through it \cite{waterman_risc-v_2021}, hence the second row in Table \ref{tab:events}), the OBC also holds a real-time counter. This register, currently, is implemented as a cycle counter of constant frequency, but without the possibility of being inhibited, thus acting as the real-time clock source for the whole core. However, the option remains that a suitable quartz crystal may be incorporated as the real-time RTC source in the future.

The next event is the execution of an instruction, also known as retirement in RISC-V terminology. This event occurs when the instruction finally completes the last stage, and the results are committed back to the register file. Up to this point, these two are the only events declared in the RISC-V specification; the rest are left platform specific.

In particular, for the platform presented in this paper, the PMU has been extended to support several other events, based on what has been deemed appropriate given the intent of this on-board processor to be for space applications.

Thus, the first additional supported events are for counting the number of exceptions and other types of traps encountered during execution, such as both timer and external interrupts. Secondly, branches, both taken and not taken, and unconditional jumps are also accounted for, triggered during the execution stage. Other instructions that merit accountability, as they often contribute to pipeline delays, are memory instructions, which generate an event for each type of operation: loads, stores and fetches. Moreover, there is a more general event to count when any instruction accesses memory through the MEM stage. Finally, hazards are monitored on the ID stage. It's worth noting that here the term `hazard' specifically refers to those instances that cannot be resolved via the forwarding unit, necessitating the insertion of a bubble in the pipeline, as the rest are transparent and do not affect the CPI.

As it can be observed, each event that could modify the cycles per instruction (CPI) metric is accounted for; hence, this way, the behavior of the OBC can be characterized as will be shown on Section \ref{sec:model}. It should be noted that not all events were supported from the start, but rather, as these tests were performed, additional events were progressively included, demonstrating the rapid and easy extensibility of the PMU.

\subsection{Implementation Details}
\label{sec:implementation}

Finally, besides the microarchitectural changes described above, to provide more depth to the modifications needed to integrate the PMU inside of an already existing RISC-V OBC, the synchronism mechanisms for counting will be explained.

First, it must be noted that according to the RISC-V specification \cite{waterman_risc-v_2019, waterman_risc-v_2021}, the PMU is defined inside the Control and Status Registers (CSRs), and hence those must be supported. These registers have the particularity of needing to be accessed atomically to prevent the formation of race conditions during the configuration of the internal state of the CPU. Thus, the PMU counters also must be accessed as such.

\begin{figure}[ht]
	\centering
	\includegraphics[width=0.5\linewidth]{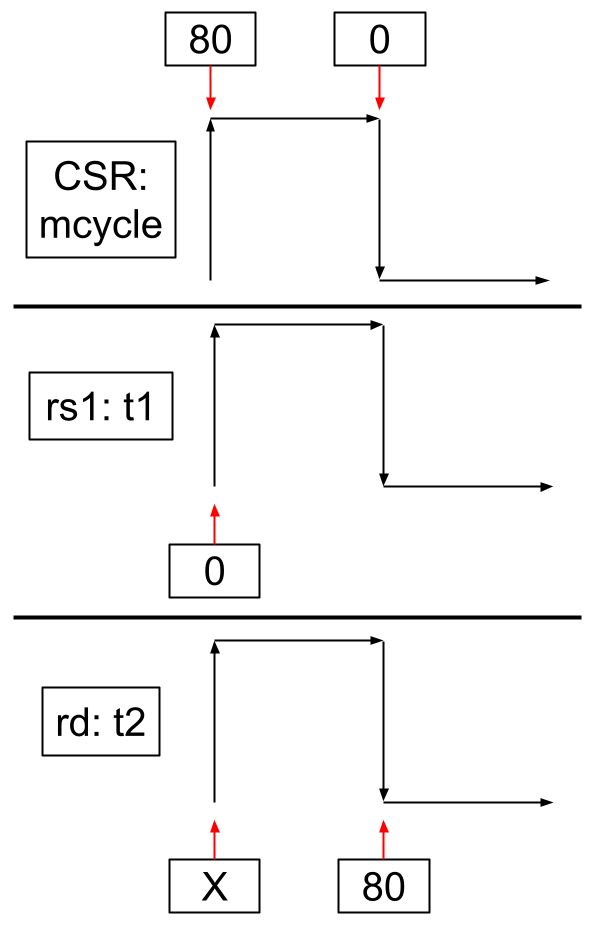}
	\caption{Example of the read and write concurrent accesses produced between CSRs and GPRs for the atomic modification of the CSRs.}
	\label{fig:csr-atomic}
\end{figure}

The selected design for atomic access consisted of performing both reads and writes in the same clock cycle, one on each clock edge. Therefore, the sequence of accesses would be the following: initially, during the rising edge, the old value of the CSR is stored on an intermediary structure at the exit of the CSR module. Later, at the falling edge, two writes occur simultaneously on different components: the corresponding CSR is written with the value of the general-purpose source register (\textit{rs1}), and the intermediate value with the former value of the CSR is stored in the destination GPR register (\textit{rd}). A visual representation of this transaction can be seen in Figure \ref{fig:csr-atomic}.

One more thing to note is that, in the case various accesses were made to the same CSR sequentially, for example with the sequence of instructions seen in Listing \ref{lst:code}, new types of data hazards need to be detected during the ID stage.

\begin{lstlisting}[language={[RISC-V]Assembler},label=lst:code,caption={Sequence of instructions that updates the \textit{mepc} CSR after a trap. These instructions would cause two new CSR hazards. First, the second instruction would need to wait until the CSR instructions finishes its execution to update the value of the \textit{t1} register in the WB stage, as the value can not be forwarded due to the atomicity of the CSRs. And second, for the same reasons, the third instruction needs to wait until the second arrives to the WB stage, to avoid a writing the wrong value to the \textit{mepc} CSR.}]
csrr t1, mepc
addi t1, t1, 4
csrw mepc, t1
\end{lstlisting}

Now, by inserting the PMU counters into the equation, there is one last additional transaction to complete during that same cycle in which a read and a write can be performed. In this instant, when the corresponding event to a specific counter has been triggered, a count must also be fulfilled. This event update also occurs on the rising edge.

However, a problem arises in this situation, since it is apparent that a write and an increment may have to be performed on the same counter. Here, 3 different scenarios might happen: first, if the write occurs in the cycle before the triggered event arrives at the count stage, the write occurs nominally and in the next cycle, once the counter is already updated after the write, the event count occurs. However, in the other two cases, if the write occurs in the same cycle or in the cycle after the count, the event will be lost, as the counter will be overwritten immediately after being counted. This is not an error but is the expected behavior, since if a counter is to be written the intention is to reset the counter, thus losing the previous count. Although all of these scenarios may appear conceptually simple, the hardware synchronization required to implement them is not trivial, due to the fact that performing the atomic accesses in the same cycle can lead to the same signal being driven with multiple inputs, generating combinational loops and other complex problems to debug. Thus, to solve this, the final design adopted the use of shadow registers: this way, after a write on the CSRs, at the falling edge, the value written would be stored in those shadow registers and the value of the CSRs would not be updated in the actual registers until the subsequent rising edge, at which point the increment would also be performed.

\subsection{Specific example of operation}
\label{sec:example}



\begin{figure*}
	\centering
	\includegraphics[width=1\linewidth]{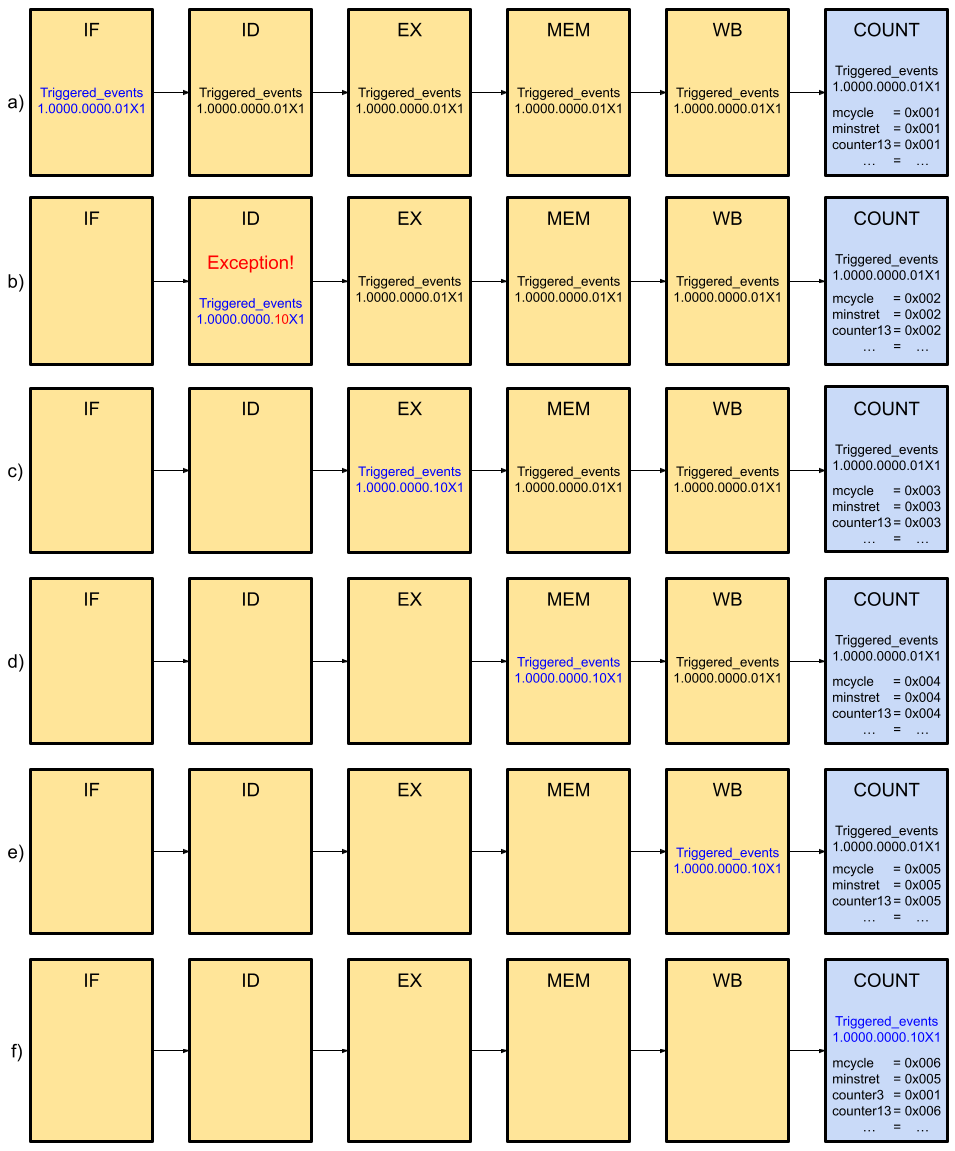}
	\caption{Example of setting and clearing of events by the PMU, and of the counting process.}
	\label{fig:example}
\end{figure*}



Now a complete step-by-step example of how the PMU operates will be presented. In Figure \ref{fig:example}, there is a visual representation of six cycles of execution of the pipeline, one in each row. As can be seen, the pipeline is full, so every cycle a new instruction is written back.

The instruction we will be paying attention to is the one marked in blue. As we can see, during the Instruction Fetch (IF) stage, subfigure \ref{fig:example}.a. everything works nominally, and, at this moment, this instruction has triggered the first, third and last events. Respectively, that is because, firstly, a cycle has already been spent by this instruction as it is always the case; secondly, at this point in time the pipeline believes the instruction is going to be retired; and thirdly, the instruction has been fetched; thus the three events triggered.

Meanwhile, to the right, in the count module, we can also see that the events of the instruction which was in the WB stage in the previous cycle are finally counted. One thing to note is that, since after the mentioned instruction was counted the value of the visible counters is 1, we must assume that the previous instruction restarted all the counters of the PMU. In addition, for the rest of this example, it can be considered that all instructions, except the one marked in blue, are nominal arithmetic and logical instructions, and therefore, only generate and count the three events mentioned previously. Furthermore, each event is accrued in its corresponding counter, as shown in Table \ref{tab:events}.

Next, in the following cycle, subfigure \ref{fig:example}.b., the instruction under analysis has arrived at the decodification stage, where it finds an exception. Since it has been detected at this stage, it could be, for example, an illegal instruction exception. This means that the instruction must not be executed and hence, the retirement event is cleared while the fourth event is set. This is because, in the proposed implementation, as also seen earlier in Table \ref{tab:events}, the fourth triggered event is programmed to count exceptions in the \textit{mhpmcounter3} register. Nevertheless, the cycle and fetch events have to be kept set, since the former must always be counted unless inhibited, and the latter because it has already happened and, thus, must be counted anyway.

In the next three cycles, subfigures \ref{fig:example}.c. - \ref{fig:example}.e., the instruction continues through the pipeline as a NOP instruction, chaining the \textit{triggered\_events} data structure through the pipeline with the same values set earlier. Finally, when it arrives at the WB stage, subfigure \ref{fig:example}.e., due to the exception it does not produce any changes to the state of the CPU. It is also at this moment, once the pipeline has been emptied, that the exception can begin to be managed, and its treatment will begin by fetching its trap handler. Finally, in the next cycle, the instruction abandons the pipeline, pending only the accounting of the events which we can see occur on the subsequent cycle, subfigure \ref{fig:example}.f.

This way we can see how, on the one hand, in the previous cycles, subfigures \ref{fig:example}.c. - \ref{fig:example}.e., the cycle, retired instruction and fetch events were counted, since they are regular instructions, as explained earlier. While, on the other hand, for the blue instruction, subfigure \ref{fig:example}.f., only cycle and fetch events are incremented due to the exception. And lastly, we can also see a new type of event, an exception, which is also counted at this point.

\section{Execution Model}
\label{sec:model}

Once the PMU was implemented it was time to test that it performed correctly. For this purpose, various pieces of software were characterized and all the events that occurred during its execution were calculated. Then, these programs were executed on the OBC and the results obtained by the PMU were observed to be precisely the ones expected. Table \ref{tab:model} shows an example of these results. In addition, the execution of these programs was followed cycle by cycle, checking that the operation was correct. These pieces of software were based on the quicksort algorithm and modified to create every type of event that the PMU is capable of detecting. More information on why this program was selected and why it is of enough significance for a space-graded on-board computer can be found in \cite{gamino_del_rio_risc-v_2020}.

Apropos, when talking about programming languages, an execution model is the way a program is processed so that each of its elements completes its determined function to achieve the final objective of the program. Hence, the two main characteristics of a programming language are its syntax, and its execution model. Then, these execution models, in some cases, can be as simple as executing each line one after the other, meanwhile in other cases, for example in General-Purpose GPU programming (GPGPU) languages, such as with CUDA \cite{farber_chapter_2011}, the use of instruction level parallelism can complicate the model of computation significantly. Another example, with a scope comparable to the use cases elaborated in this paper is \cite{pellizzoni_predictable_2011}, where a discussion on an execution model for real-time embedded languages can be found.

Applying the same concept on a lower abstraction layer, analogously, it can be inferred that every CPU has its own execution model. In modern general-purpose CPUs, for instance, they can be so complex that simulating and obtaining a WCET is extremely difficult without big margin errors. This is disfavored in space applications and thus it is common to use simpler CPUs with more deterministic models.

Obtaining the execution model of a CPU can be extremely useful to then be capable of simulating and predicting its behavior, e.g., with the use of cycle-accurate simulators \cite{sanchez_arinc653_2021}. This results in a reduction of the development time and costs for both software and hardware, and an ease in the complexity of the debugging process, for instance, when adding new functionality to the CPU, to check whether errors or other unintended side effects have been introduced.  

For these reasons, after having validated the functionality of the PMU with the tests described earlier, all the potential events which altered the CPI of the processor and their corresponding amount of time spent executing were already available, as mentioned on Section \ref{sec:configuration}. Therefore, with this information, the current execution model has been calculated, which characterizes the behavior of the presented OBC.

Now this model will be described. For the sake of brevity, as this is not the main topic of the article, but simply a way to provide insight into the advantages gained by integrating the PMU, the explanations of the execution model will not be elaborated in depth, but rather merely an overview of its intricacies will be provided.

For any sequential program, i.e., without instructions that change the control flow, the total number of cycles equals the number of instructions executed, the number of instructions fetched, and the number of hazards found during execution plus 4. These 4 cycles are due to the pipeline being filled at the start of execution and must always be accounted for. Moreover, another thing to consider is that memory access instructions take longer to execute than any other type of instruction, as there is a latency between the petition and receiving the respective data or acknowledgment from memory. In these cases, stores take one extra cycle and loads take two extra cycles to talk to memory, i.e., in total, each of these instructions takes two and three cycles, respectively, to end, with one of them dedicated to the normal propagation to the next stage of the pipeline.

Adding to the complexity, the next case is when a jump or branch instruction is executed. In those cases, 2 cycles must be added to the previous execution model for each of these instructions executed since the CPU commits the jump during the EX-stage and therefore, 2 cycles are lost filling the pipeline again.

Finally, when any trap is encountered (either interrupts or exceptions), they must be considered like a flush and its subsequent refilling of the pipeline. Hence, for both each entry and exit to the trap, 4 cycles are to be accounted for. This is because, on entry, the fetch of the trap handler is not carried out until the pipeline has been completely emptied (as another type of trap could be encountered during the emptying process and therefore take precedence). While, during the MRET execution for the trap handler exit, the pipeline also needs to be emptied to make sure that no instruction previous to it is modifying the \textit{mepc} CSR, to know where to resume the execution, and to make sure all instructions are executed with the appropriate privilege level. There is an exception to this rule when a MRET is immediately followed by another trap, in which case a cycle is lost, and only 7 cycles (instead of 8) are taken for an entire trap.

With all of this information in mind, as mentioned earlier, a program can be characterized with the number of events that would happen during its execution and then checked empirically by hand. Then, with the execution model of the OBC, it can be observed, for example, that the number of cycles counted by the PMU is the same as the number of cycles calculated theoretically to be the length of execution of such a program. Thus, this way, the behavior of the PMU and the OBC can be validated. An example of this can be seen in Table \ref{tab:model}.

\begin{table}[h]
\centering
\resizebox{\linewidth}{!}{
\begin{tabular}{|c|c|c|c|}
\hline
Event                	& \begin{tabular}[c]{@{}c@{}}Event \\ count\end{tabular} & \begin{tabular}[c]{@{}c@{}}Cycles \\ per event\end{tabular} & \begin{tabular}[c]{@{}c@{}}Total cycles \\ per event\end{tabular} \\ \hline
Cycles               	& 119540                                             	& -                                                       	& -                                                             	\\ \hline
Retired instructions 	& 32950                                              	& 1                                                       	& 32950                                                         	\\ \hline
Exceptions           	& 0                                                  	& 8                                                       	& 32950                                                         	\\ \hline
External interrupts  	& 0                                                  	& 8                                                       	& 32950                                                         	\\ \hline
Timer interrupts     	& 0                                                  	& 8                                                       	& 32950                                                         	\\ \hline
Branches             	& 1396                                               	& 2                                                       	& 35742                                                         	\\ \hline
Jumps                	& 1380                                               	& 2                                                       	& 38502                                                         	\\ \hline
Hazards              	& 9519                                               	& 1                                                       	& 48021                                                         	\\ \hline
Loads                	& 13667                                              	& 2                                                       	& 75355                                                         	\\ \hline
Stores               	& 5675                                               	& 1                                                       	& 81030                                                         	\\ \hline
Fetches              	& 38506                                              	& 1                                                       	& 119536                                                        	\\ \hline
Initial pipeline filling & -                                                  	& 4                                                       	& 119540                                                        	\\ \hline
\end{tabular}
}
\caption{Characterization of a quicksort program with 64 elements. In the first column are all the events that modify the CPI during the execution of the program, as has already been explained, whereas, in the second column are the number of counts of each event, as obtained by the PMU after execution. In the third column, it can be seen the amount of cycles spent per each event due to the OBC execution model, as has been discussed. And finally, in the fourth column are the total cycles employed during the program due to each event. The bottom rightmost cell shows how the theoretical number of cycles obtained from the execution model coincides with the number of cycles measured by the PMU (in the topmost cell of the second column).}
\label{tab:model}
\end{table}

\section{Experimental Results}
\label{sec:results}

Once the behavior of the PMU has been validated, both with the first tests discussed above as well as with the execution model, as a real-world showcase of its use, the results obtained during the execution of several programs will be exhibited. However, to ensure completeness, for the purpose of this paper, the configuration of the PMU used for the tests will be described first.

\begin{table}[htb]
\centering
\resizebox{\columnwidth}{!}{
\begin{tabular}{|c|c|c|}
\hline
Configuration Register & Values                	\\ \hline
mcounteren         	& 0xFFFFFFFF            	\\ \hline
mcountinhibit      	& 0x000000000 or 0xFFFFFFFF \\ \hline
\end{tabular}
}
\caption{General configuration registers.}
\label{tab:gen-config-CSRs}
\end{table}

The values of the general configuration registers have been what can be seen in Table \ref{tab:gen-config-CSRs}. First, with \textit{mcounteren}, the PMU counters have been set up to also be accessible through user mode to reduce the number of calls to the execution environment and to ease the software development. Secondly, with \textit{mcountinhibit}, the counters were inhibited both during the initial setup and when it was time to read them, to be able to gather the statistics atomically, whereas, the rest of the time, they were uninhibited. Moreover, in regards to the specific configuration registers and which events were programmed to be monitored, they were set up according to the order of Table \ref{tab:events}, as explained in the description in Section \ref{sec:configuration}.

Throughout the remainder of this section, both the performance metrics and the statistical results obtained by the PMU during the execution of different benchmarks will be exhibited. Furthermore, a comparative analysis between the results of the presented OBC and those of another processor is also conducted. Finally, the resource utilization and power consumption data are analyzed to provide a comprehensive view of the operating efficiency and power consumption details of the processor after the implemented modifications.

\subsection{Benchmarks performance results}
\label{sec:benchmarks}

Before reviewing the results, it should be noted that the performance metrics shown below are with the PMU enabled, since without it, measuring the performance of the OBC would be much more complicated, as this is precisely what we want to achieve with its integration. It could have been possible with an external real-time clock synchronized with the start and end of the execution of each program. However, given the bare-metal nature of the platform, which would make such synchronization rather difficult, and the nanosecond variations that would be encountered at most, it was considered unnecessary since the PMU, by design, does not modify performance. This is because, as described in Section \ref{sec:design}, the PMU is not intrusive with respect to execution, as it does not affect the way instructions are executed since it only performs its monitoring role, nor does it affect the duration of instructions in the pipeline, since they continue to be written back during the same stage as before the modification. Thus, the results shown are analogous to those without PMU enabled.

Specifically, at the moment of publication, both Dhrystone and CoreMark have been ported to our platform, as they are the most common benchmarks in the industry. Another reason for using them was that RISC-V versions of these benchmarks already existed, such as those in the NEORV32 processor Board Support Packages (BSPs) \cite{nolting_neorv32_2022}, which significantly eased the porting process. Both benchmarks were compiled with the RISC-V GNU toolchain \cite{gcc_risc-v_2023}.

\begin{figure*}[bp]
\centering
\begin{subfigure}{0.495\textwidth}
	\includegraphics[width=\textwidth]{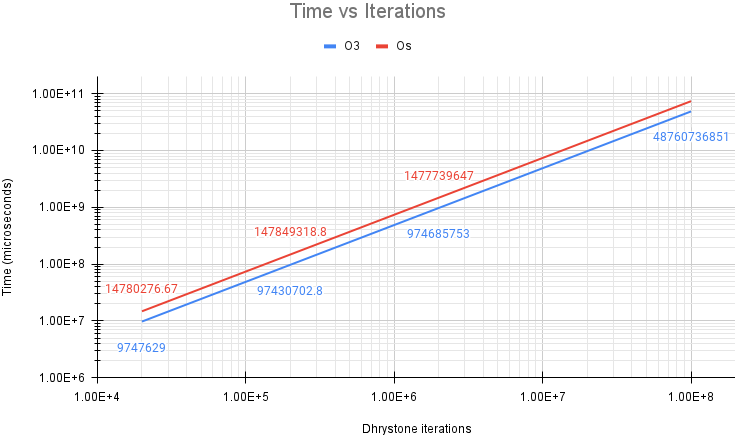}
	\caption{}
	\label{fig:time-iterations-dhrystone}
\end{subfigure}
\hfill
\begin{subfigure}{0.495\textwidth}
	\includegraphics[width=\textwidth]{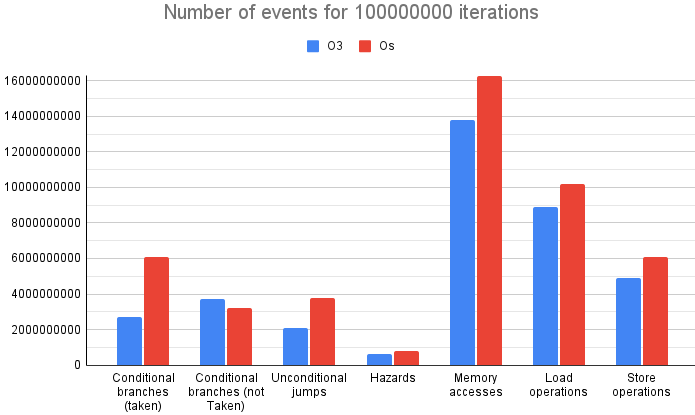}
	\caption{}
	\label{fig:events-dhrystone}
\end{subfigure}

\begin{subfigure}{0.495\textwidth}
\vspace{1em}
	\includegraphics[width=\textwidth]{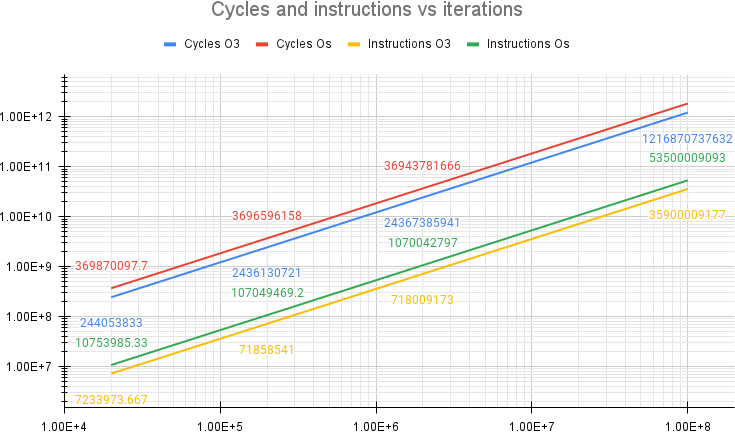}
	\caption{}
	\label{fig:cycles-instructions-iterations-dhrystone}
\end{subfigure}
\hfill
\begin{subfigure}{0.495\textwidth}
	\includegraphics[width=\textwidth]{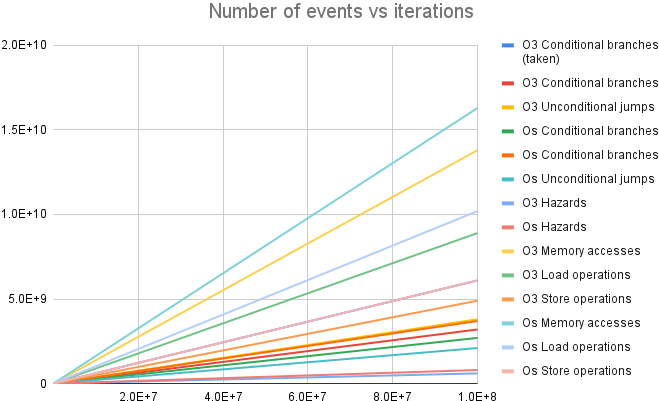}
	\caption{}
	\label{fig:events-iterations-dhrystone}
\end{subfigure}
   	 
\caption{Results of the Dhrystone benchmark. Subfigures \ref{fig:time-iterations-dhrystone}, \ref{fig:cycles-instructions-iterations-dhrystone} and \ref{fig:events-iterations-dhrystone} show the linear relationship for each metric over the entire number of iterations. Specifically, the real-time spent executing, cycles and instructions executed and all the other supported events, respectively. It should be noted that the axes are in logarithmic scale. Meanwhile, subfigure \ref{fig:events-dhrystone} exhibits the count of each event that occurred on tests with 1000000000 iterations.}
\label{fig:dhrystone}
\end{figure*}

\begin{figure*}
\centering
\begin{subfigure}{0.495\textwidth}
	\includegraphics[width=\textwidth]{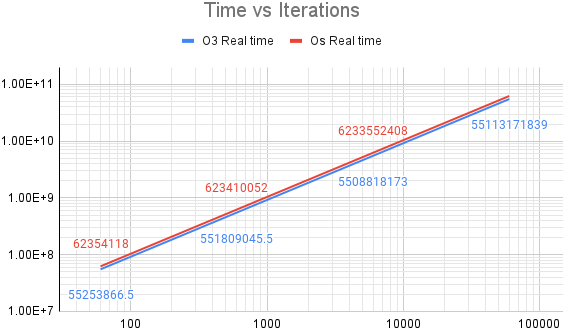}
	\caption{}
	\label{fig:time-iterations-coremark}
\end{subfigure}
\hfill
\begin{subfigure}{0.495\textwidth}
	\includegraphics[width=\textwidth]{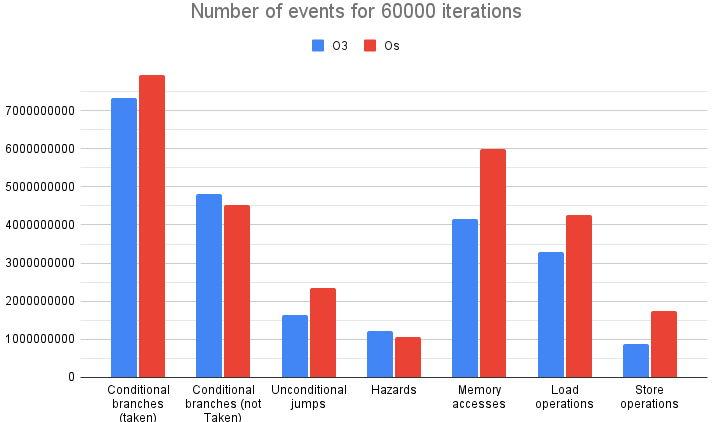}
	\caption{}
	\label{fig:events-coremark}
\end{subfigure}

\begin{subfigure}{0.495\textwidth}
\vspace{1em}
	\includegraphics[width=\textwidth]{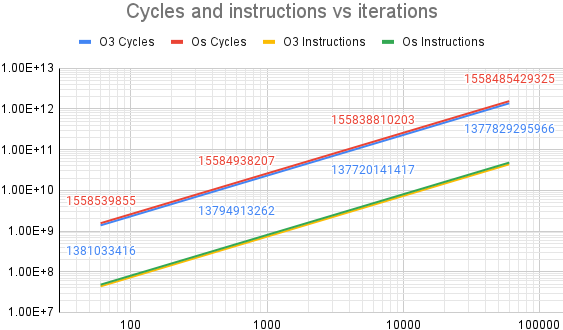}
	\caption{}
	\label{fig:cycles-instructions-iterations-coremark}
\end{subfigure}
\hfill
\begin{subfigure}{0.495\textwidth}
	\includegraphics[width=\textwidth]{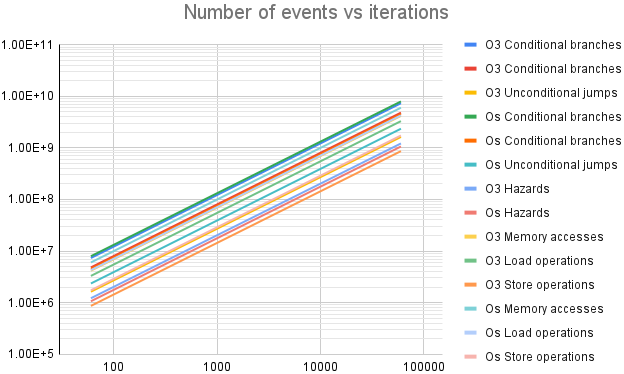}
	\caption{}
	\label{fig:events-iterations-coremark}
\end{subfigure}
   	 
\caption{Results of the CoreMark benchmark. Subfigures \ref{fig:time-iterations-coremark}, \ref{fig:cycles-instructions-iterations-coremark} and \ref{fig:events-iterations-coremark}  show the linear relationship for each metric over the entire number of iterations. Specifically, the real-time spent executing, cycles and instructions executed and all the other supported events, respectively. It should be noted that the axes are in logarithmic scale. Meanwhile, subfigure \ref{fig:events-coremark} exhibits the count of each event that occurred on tests with 60,000 iterations.}
\label{fig:coremark}
\end{figure*}

Once the porting process was completed, the benchmarks were ready to be executed. Instead of simply providing a single data point with a fixed number of iterations, the benchmarks were left running for different lengths of time and with different compiler optimizations to examine the consistency of the results. This approach enabled not only performance evaluation but also to test the temporal consistency of the processor, and thus validate the processor's deterministic and linear behavior. It is important to note that, when tests with the same number of iterations were conducted, the results were identical across all experiments. These tests were successfully conducted on various computers with multiple evaluation boards to verify the reproducibility of the results. The test environments consisted of the Vivado HDL Suite version 2018.3 \cite{amd_xilinx_2018} used to program two different Nexys4-DDR evaluation boards \cite{digilent_nexys_2016}. The summary of all executions can be seen in Figures \ref{fig:dhrystone} and \ref{fig:coremark} respectively for each benchmark. Meanwhile, the performance results for each number of iterations tested are in Tables \ref{tab:dhrystone} and \ref{tab:coremark}.

\begin{table*}
\centering
\resizebox{\linewidth}{!}{
\begin{tabular}{|c|c|c|c|c|c|c|c|c|}
\hline
Optimization                 	& Os        	& Os      	& Os     	& Os    	& O3        	& O3      	& O3     	& O3    	\\ \hline
Iterations                   	& 100000000 	& 2000000 	& 200000 	& 20000 	& 100000000 	& 2000000 	& 200000 	& 20000 	\\ \hline
Real Time ($\mu s$)               	& 73852000010   & 1477739647  & 147849319  & 14780277  & 48760736851   & 974685753   & 97430703   & 9747629   \\ \hline
Cycles                       	& 1846300363478 & 36943781666 & 3696596158 & 369870098 & 1216870737632 & 24367385941 & 2436130721 & 244053833 \\ \hline
Instructions                 	& 53500009093   & 1070042797  & 107049469  & 10753985  & 35900009177   & 718009173   & 71858541   & 7233974   \\ \hline
CPI                          	& 34.5103   	& 34.5255 	& 34.5317	& 34.3938   & 33.8961   	& 33.9374 	& 33.9018	& 33.7372   \\ \hline
Exceptions                   	& 0         	& 0       	& 0      	& 0     	& 0         	& 0       	& 0      	& 0     	\\ \hline
External interrupts:         	& 0         	& 0       	& 0      	& 0     	& 0         	& 0       	& 0      	& 0     	\\ \hline
Timer interrupts             	& 0         	& 0       	& 0      	& 0     	& 0         	& 0       	& 0      	& 0     	\\ \hline
Conditional branches (taken) 	& 6100002926	& 122002925   & 12202923   & 1222924   & 2700002952	& 54002950.67 & 5402949	& 542948	\\ \hline
Conditional branches (not Taken) & 3200000038	& 64000038	& 6400038	& 640038	& 3700000063	& 74000063	& 7400063	& 740063	\\ \hline
Unconditional jumps          	& 3800000093	& 76000093	& 7600093	& 760093	& 2100000027	& 42000027	& 4200027	& 420027	\\ \hline
Hazards                      	& 800002913 	& 16002912	& 1602910	& 162910	& 600002955.5   & 12002954.67 & 1202953	& 122952	\\ \hline
Memory accesses              	& 16300003000   & 326002999   & 32602997   & 3262997   & 13800003064   & 276003063.5 & 27603061   & 2763059.5 \\ \hline
Load operations              	& 10200002935   & 204002934   & 20402932   & 2042931   & 8900002999	& 178002997.7 & 17802996   & 1782995   \\ \hline
Store operations             	& 6100000065	& 122000065   & 12200065   & 1220064   & 4900000065	& 98000065	& 9800065	& 980065	\\ \hline
Dhrystones ($DMIPS/s$)         	& 1354.0595 	& 1353.4192   & 1352.7307  & 1353.1558 & 2050.8358 	& 2051.9482   & 2052.7444  & 2051.7855 \\ \hline
VAX $DMIPS/s$                  	& 0.7707    	& 0.7703  	& 0.7699 	& 0.7702	& 1.1672    	& 1.1679  	& 1.1683 	& 1.1678	\\ \hline
$\mu s$ per run of Dhrystone      	& 738.5200  	& 738.8698	& 739.2466   & 739.0138  & 487.6074  	& 487.3429	& 487.1535   & 487.3815  \\ \hline
\end{tabular}
}
\caption{Performance results of the Dhrystone benchmark.}
\label{tab:dhrystone}
\end{table*}

\begin{table*}
\centering
\resizebox{\linewidth}{!}{
\begin{tabular}{|c|c|c|c|c|c|c|c|c|}
\hline
Optimization                 	& Os        	& Os       	& Os      	& Os      	& O3        	& O3       	& O3      	& O3      	\\ \hline
Iterations                   	& 60000     	& 6000     	& 600     	& 60      	& 60000     	& 6000     	& 600     	& 60      	\\ \hline
Real Time ($\mu s$)               	& 62339417173   & 6233552408   & 623410052   & 62354118	& 55113171839   & 5508818173   & 551809045.5 & 55253866.5  \\ \hline
Cycles                       	& 1558485429325 & 155838810203 & 15584938207 & 1558539855  & 1377829295966 & 137720141417 & 13794913262 & 1381033416  \\ \hline
Instructions                 	& 48134147340   & 4813416078   & 481341900   & 48135534 	& 43721970101   & 4372198097   & 437220062   & 43723094	\\ \hline
CPI                          	& 32.37795859   & 32.37592754  & 32.3781047  & 32.37816125 & 31.51343118   & 31.49906257  & 31.55141875 & 31.58590322 \\ \hline
Exceptions                   	& 0         	& 0        	& 0       	& 0       	& 0         	& 0        	& 0       	& 0       	\\ \hline
External interrupts:         	& 0         	& 0        	& 0       	& 0       	& 0         	& 0        	& 0       	& 0       	\\ \hline
Timer interrupts             	& 0         	& 0        	& 0       	& 0       	& 0         	& 0        	& 0       	& 0       	\\ \hline
Conditional branches (taken)     & 7931315815    & 793131617    & 79313245    & 7931360     & 7331892933    & 733189424    & 73319043    & 7332035     \\ \hline
Conditional branches (not Taken)& 4529380830    & 452938073    & 45293730    & 4529363     & 4813800687    & 481379947    & 48137876    & 4813666     \\ \hline
Unconditional jumps     		 & 2348400587    & 234840174    & 23484047    & 2348520     & 1627148804    & 162715021    & 16271562    & 1627297     \\ \hline
Hazards                      	& 1062969079	& 106296958	& 10629769	& 1063027 	& 1216686573	& 121668690	& 12166892	& 1216722 	\\ \hline
Memory accesses              	& 5999117629	& 599912373	& 59991349	& 5999745 	& 4157900927	& 415790466	& 41579202	& 4158294 	\\ \hline
Load operations              	& 4264161951	& 426416559	& 42641751	& 4264539 	& 3294413599	& 329441606	& 32944278	& 3294674 	\\ \hline
Store operations             	& 1734955678	& 173495814	& 17349598	& 1735206 	& 863487328 	& 86348860 	& 8634924 	& 863620  	\\ \hline
Coremarks                    	& 0.9625    	& 0.9625   	& 0.9624  	& 0.9622  	& 1.0887    	& 1.0892   	& 1.0873  	& 1.0859  	\\ \hline
\end{tabular}
}
\caption{Performance results of the Coremark benchmark.}
\label{tab:coremark}
\end{table*}

As can be seen in these results, the execution time, along with the number of cycles and instructions executed, and every other type of supported event, increases linearly with the amount of iterations of each benchmark, fulfilling the established validation objective. This is due to the fact that, as explained in Section \ref{sec:design}, these tests were run without cache. Moreover, it is noteworthy that identical results were obtained across multiple tests for each iteration count, reinforcing the robustness of the findings.

Also, in Tables \ref{tab:dhrystone} and \ref{tab:coremark}, the CPI and other performance metrics of both Dhrystone and CoreMark can be observed. In order to understand these results, it must be mentioned again that performance has not been a priority during the development of this OBC; instead, the objective has always been to create a test bed to work on different proofs of concept of novel tools to increase the reliability of the hardware and software developed, as discussed in previous sections. Nonetheless, some performance improvements are already planned for the future.

In addition, the results are very sensitive to the optimization level used during the compilation process, as is common with these kinds of benchmarks. For example, the execution time is significantly lower on the tests compiled with the ``-O3'' flag instead of optimizing for program size ``-Os''. Moreover, the distribution of the remaining events shifts remarkably depending on the optimization flags, as can be seen in Subfigures \ref{fig:events-dhrystone} and \ref{fig:events-coremark}. For instance, as also seen in Subfigures \ref{fig:cycles-instructions-iterations-dhrystone} and \ref{fig:cycles-instructions-iterations-coremark}, the number of instructions, especially jumps (both conditional and unconditional), and memory access operations, are significantly lower in the versions compiled with the ``-O3'' flag, demonstrating the great predictive and optimization capabilities of existing compilers. This is especially notable in the Dhrystone benchmark, with taken and unconditional jumps reducing nearly in half, whereas in CoreMark, although these differences still exist, they are much more succinct. Of special significance is the \textit{-fpredictive-commoning} optimization, which reuses computations made during the flow of a program, most notably memory loads and stores performed in previous iterations of loops, thus the variations observed in the figures. These disparities explain most of the temporal deviation between execution times.

To try and mitigate the discrepancies created with software optimizations, the addition of newer and improved benchmarks such as Embench \cite{bennett_embench_2019} is a work in progress for the foreseeable future. Furthermore, a port to our RISC-V platform of the Boot Software of the Instrument Control Unit (ICU) of the Energetic Particle Detector (EPD) currently on-board the Solar Orbiter spacecraft is already ongoing at the moment of publication.

Finally, it is important to note that the PMU has already proven extremely useful during our workflow, as it has improved the development of both our hardware and our software, facilitating the debugging of issues during the OBC design and implementation and adding reliability to our tests.

\subsection{Comparison with other processors}
\label{sec:comparison}

In this subsection, a comparative analysis is conducted between the statistical results obtained by executing two different sets of tests on the presented OBC and an alternative processor. This is because, although with the results showcased in section \ref{sec:model} the behavior of the proposed CPU and the PMU had already been internally validated, its comparison with another processor would help validate them against an external reference. Moreover, such comparison would help further illustrate the impact achieved with the modifications proposed in the design of this PMU.

The processor selected for the comparison has been the NEORV32 \cite{nolting_neorv32_2022}. Its choice is informed by its prominence within the RISC-V community, its alignment with our area of focus, as even though it is not specifically targeted for space applications, it is also designed as a microcontroller, and our existing familiarity with its architecture, as discussed in the preceding subsection. Additionally, despite challenges  stemming from differences in event coverage arising from distinct pipeline structures, as its architecture is not fully pipelined, key metrics such as the number of instructions executed, jumps, and memory operations still enable meaningful comparison. 

The first set of tests can be found in Table \ref{tab:comparison_coremark}. In it, the results of executing the Coremark benchmark with the same configuration as in the previous subsection on both processors can be found. In the interest of clarity and conciseness, these analyses focus solely on the outcomes derived from the Coremark benchmark with the -Os compiler optimization flag. While the preceding section encompassed tests for both Coremark and Dhrystone benchmarks, including results from both -Os and -O3 compiler optimization flags, ultimately, the same findings were reached for all tests. Consequently, it has been concluded that the results obtained with the Coremark benchmark under the -Os optimization flag adequately represent the performance characteristics of the processors under study. Thus, to avoid redundancy and maintain relevance, the decision has been made not to duplicate the analysis by including additional result sets in this comparison.

It is important to acknowledge there are some significant disparities between the results presented in this section and those in the preceding one. These differences arise from the adjustments made to the memory interface of the proposed OBC to ensure closer comparability with the NEORV32. Unlike the NEORV32, which operates solely with integrated memory, the results shown in \ref{sec:benchmarks} utilize the external memory provided by the Nexys4-DDR development board \cite{digilent_nexys_2016}. This approach was chosen for its broader applicability, reproducibility, and readiness for future expansions. However, to ease the comprehension of the comparison with the NEORV32, modifications were made to the memory access mechanism. Specifically, rather than utilizing the Memory Interface Generator (MIG) controller for external memory access, the system’s main memory was synthesized within the FPGA. This alteration significantly reduces latency in accesses to memory, leading to improved performance as observed in the respective columns for the presented OBC in Table \ref{tab:comparison_coremark}. Nevertheless, it can be seen that despite these performance enhancements, the remaining statistical results remain unchanged.

Upon examination of Table \ref{tab:comparison_coremark}, it is evident that the statistics obtained with the PMU of each processor are consistent with each other. Firstly, the number of instructions precisely matches for each individual set of tests, considering even the executions featuring vastly different numbers of iterations. Secondly, while the NEORV32 lacks events to differentiate between each type of branch instruction, the results for taken and total branch instructions align precisely with the totals obtained in the proposed PMU. Lastly, even though the number of cycles and the real-time duration of execution are inherently non-comparable, due to differences in the microarchitectures, their results remain coherent. The proposed processor, featuring a pipelined architecture, exhibits a lower CPI compared to the NEORV32's multicycle architecture with pipelined fetch and execute stages, resulting in slightly longer execution times for the latter in each test. The congruence displayed across these tests underscores the reliability and accuracy of the measurements from the proposed PMU.

\begin{table*}
\centering
\resizebox{\linewidth}{!}{
\begin{tabular}{|c|c|c|c|c|c|c|c|c|}
\hline
Processor   		         	& RV32Xtrace   & NEORV32  	& RV32Xtrace	& NEORV32   	   & RV32Xtrace	& NEORV32  	   & RV32Xtrace	& NEORV32  	 \\ \hline
Optimization   			 	& Os  			& Os  		   & Os  		   & Os    	   	& Os  		   & Os 		   & Os    	   	& Os    		 \\ \hline
Iterations                  	& 60000   		& 60000    	& 6000   	   & 6000    	   & 600   	        & 600 		   & 60   	   	& 60    		 \\ \hline
Real Time ($\mu s$)    		 & 5777969880   & 7939301740   & 577797146      & 793930514 	& 57779763  	& 79393149       & 5778134       & 7939655     \\ \hline
Cycles             			 & 144449246994 & 198482543494 & 14444928651   & 19848262843   & 1444494086      & 1984828714	& 144453360 	& 198491365   \\ \hline
Instructions       			 & 48134147340  & 48134147340  & 4813416078	& 4813416078	& 481341900       & 481341900      & 48135534      & 48135534  	 \\ \hline
CPI                			 & 3,0009723    & 4,1235288    & 3,0009723     & 4,1235294     & 3,0009731     & 4,1235319     & 3,0009713     & 4,1235933   \\ \hline
Exceptions         			 & 0        	& 0  	   	& 0 		   	& 0 		   	& 0   		   & 0  		   & 0 		   	& 0 			 \\ \hline
External interrupts:   		 & 0        	& 0   		   & 0 		   	& 0 		   	& 0   		   & 0  		   & 0 		   	& 0 			 \\ \hline
Timer interrupts   			 & 0        	& 0  		   & 0 		   	& 0 		   	& 0   		   & 0  		   & 0 		   	& 0 			 \\ \hline
Conditional branches (taken)	& 7931315815   & n.a.		   & 793131617 	& n.a.   	   & 79313245      & n.a.  		   & 7931360   	& n.a.   	 \\ \hline
Unconditional jumps         	& 2348400587   & n.a.		   & 234840174 	& n.a.   	   & 23484047       & n.a.  		   & 2348520   	& n.a.   	 \\ \hline
Total jumps taken      		 & 10279716402  & 10279716402  & 1027971791	& 1027971791	& 102797292 	& 102797292 	& 10279880  	& 10279880    \\ \hline
Conditional branches (not Taken)& 4529380830   & n.a.		   & 452938073 	& n.a.   	   & 45293730 	   & n.a.    	   & 4529363   	& n.a.   	 \\ \hline
Total jumps                 	& 14809097232  & 14809097232  & 1480909864	& 1480909864	& 148091022 	& 148091022 	& 14809243  	& 14809243  	 \\ \hline
Hazards            			 & 1062969079   & n.a.		   & 106296958 	& n.a.   	   & 10629769      & n.a.  		   & 1063027   	& n.a.   	 \\ \hline
Memory accesses             	& 5999117629   & n.a.		   & 599912373 	& n.a.   	   & 59991349       & n.a.  		   & 5999745   	& n.a.   	 \\ \hline
Load operations    			 & 4264161951   & 4264161951   & 426416559 	& 426416559 	& 42641751  	   & 42641751  	   & 4264539   	& 4264539   	 \\ \hline
Store operations   			 & 1734955678   & 1734955678   & 173495814 	& 173495814 	& 17349598  	& 17349598  	& 1735206   	& 1735206   	 \\ \hline
Coremarks          			 & 10,384270123 & 7,557339670  & 10,384267284  & 7,557336434   & 10,384258585  & 7,557327144   & 10,38397517   & 7,557003421 \\ \hline
\end{tabular}
}
\caption{Comparison between the results from executing the Coremark benchmark on each CPU for each number of iterations.}
\label{tab:comparison_coremark}
\end{table*}

Furthermore, to better illustrate the differences in processor count influenced by the proposed PMU design in more intricate scenarios, we introduce Table \ref{tab:comparison_exception}. This table offers a comparative analysis wherein an exception is encountered every $100000$ instructions. The objective of this set of tests is to scrutinize the impact of the proposed PMU design on its statistical outcomes in more real-world scenarios where traps are commonplace, rather than only through benchmarks with nominal runs. These tests were also configured to execute for different amounts of time, to evaluate the temporal consistency and reproducibility of the results.

In contrast to the previous set of tests, this analysis poses greater complexity, requiring an understanding of how each processor establishes its counting strategies. For instance, in the NEORV32, all control flow transfer operations are collectively tallied within a single counter, labeled `Total jumps taken’ in Table \ref{tab:comparison_exception}. This counter includes all branch instructions, both conditional and unconditional, as well as trap entries and exits. However, in the event named `Total jumps instructions’, only the jumps due to instructions are accounted for, both taken and not taken. In contraposition, the proposed design distinguishes each type of branch instruction and assumes the occurrence of an entry and exit jump for every trap, as they must necessarily always be realized. Therefore, the NEORV32 appears to register two additional branches for each trap raised within the former counter, a divergence absent in the latter counter. Thus, these discrepancies emerge from counting the entry and exit jumps for each trap within that same first counter, whereas the latter counter does indeed track the same events, resulting in identical counts. Consequently, the counts for these events are equivalent, with the disparities arising solely from the counting strategy of each processor. Meanwhile, the counts for the remaining events tracked in both processors match exactly.

Nevertheless, it is evident that the NEORV32 processor registers a greater total count of instructions executed compared to the proposed design, with the increment directly correlating to the number of encountered traps. This disparity underscores the limitation discussed throughout this article, which serves as the motivation for the proposed design. Unlike the NEORV32, which lacks the capability to retract an event once triggered, the proposed design aims to address this issue, ensuring a more reliable count.

\begin{table*}
\centering
\resizebox{\linewidth}{!}{
\begin{tabular}{|c|c|c|c|c|c|c|c|c|}
\hline
Processor         			 	& RV32Xtrace & NEORV32  & RV32Xtrace & NEORV32	& RV32Xtrace   & NEORV32  	\\ \hline
Optimization      			 	& Os		  & Os   	& Os     	& Os     	& Os       	& Os       	\\ \hline
Iterations        			 	& 1 		  & 1    	& 10000000   & 10000000   & 1000000000   & 1000000000   \\ \hline
Real Time ($\mu s$)   		 	& 12     	& 16   	& 100208622  & 131524896  & 17445198786  & 23229902382  \\ \hline
Cycles            			 	& 305    	& 404  	& 2505215566 & 3288122409 & 436129969670 & 580747559552 \\ \hline
Instructions      			 	& 91     	& 92   	& 801604662  & 801604762  & 139815669906 & 139815679906 \\ \hline
CPI               			 	& 3,329670   & 4,391304 & 3,1252507  & 4,1019247  & 3,119321 	& 4,153665 	\\ \hline
Exceptions        			 	& 1      	& 1    	& 100    	& 100    	& 10000    	& 10000    	\\ \hline
External interrupts:  		 	& 0 		  & 0    	& 0      	& 0      	& 0        	& 0        	\\ \hline
Timer interrupts  			 	& 0 		  & 0    	& 0      	& 0      	& 0        	& 0        	\\ \hline
Conditional branches (taken)	& 4      	& n.a. 	& 160499981  & n.a.   	& 28329532664  & n.a.     	\\ \hline
Unconditional jumps   		 	& 8      	& n.a. 	& 50000102   & n.a.   	& 9294977298   & n.a.     	\\ \hline
Total jumps taken 			 	& 12     	& 14   	& 210500083  & 210500283  & 37624509962  & 37624529962  \\ \hline
Conditional branches (not Taken)& 7      	& n.a. 	& 148700093  & n.a.   	& 24822309986  & n.a.     	\\ \hline
Total jump instructions            	 	& 19     	& 19   	& 359200176  & 359200176  & 62446819948  & 62446819948  \\ \hline
Hazards           			 	& 14     	& n.a. 	& 20000705   & n.a.   	& 2000070005   & n.a.     	\\ \hline
Memory accesses   			 	& 46     	& n.a. 	& 40003605   & n.a.   	& 4000360005   & n.a.     	\\ \hline
Load operations   			 	& 25     	& 25   	& 30001804   & 30001804   & 3000180004   & 3000180004   \\ \hline
Store operations  			 	& 20     	& 20   	& 10001801   & 10001801   & 1000180001   & 1000180001   \\ \hline
\end{tabular}
}
\caption{Comparison between the statistical results on each processor after the execution of the different tests with exceptions.}
\label{tab:comparison_exception}
\end{table*}

Lastly, it is important to note that the example portrayed here represents a relatively mild scenario in comparison with the severe examples mentioned in Section \ref{sec:design} when the motivation of this design was presented. This is because, as briefly mentioned earlier, the NEORV32 possesses a multicycle architecture and thereby only a single instruction is executed at a time. The rationale for this design is also to be able to provide precise trap control \cite{nolting_neorv32_2024-1}. Therefore, as there is only one instruction executing, when a trap arrives, it is easier to control which events need to be reverted in comparison with a pipelined architecture, where multiple instructions are executing concurrently. Hence, it is reasonable to infer that processors with less emphasis on performing precise and accurate counting in trap situations, may be even more significantly impacted in this scenario.

\subsection{Utilization and power results}
\label{sec:utilization_power}

Once the performance results have been described, now the resource and power utilization will be portrayed. These results are particularly important in this case because of the field of application of this on-board computer, due to the significant power and mass limitations in the space industry, as it is well known.

With regard to resource utilization, Tables \ref{tab:utilizationPMU} and \ref{tab:utilization-without-PMU} present a summary of the main resources used by the OBC, both with and without the PMU enabled, respectively. All of these metrics were extracted from the Vivado Suite analysis report tool and the results provided are those obtained during synthesis with exactly the same settings selected for both projects.

As can be imagined, most of the extra resources used by the design with PMU are in the CSR unit, due to the new registers and logic implemented; being the remaining extra resources allocated in the inter-stage registers where the events are monitored and then propagated through the remaining pipeline, with the \textit{triggered\_events} data structure.

\begin{table}[htb]
\centering
\resizebox{0.95\linewidth}{!}{
\begin{tabular}{|c|c|c|c|c|}
\hline
Component   	& Slice LUTs & Slice Registers & F7 Muxes & Block RAM \\ \hline
RV32        	& 6498   	& 5347        	& 292  	& 2     	\\ \hline
clint       	& 77     	& 72          	& 0    	& 0     	\\ \hline
decode      	& 334    	& 44          	& 0    	& 0     	\\ \hline
execute     	& 634    	& 0           	& 1    	& 0     	\\ \hline
fu          	& 74     	& 0           	& 0    	& 0     	\\ \hline
if          	& 22     	& 35          	& 0    	& 0     	\\ \hline
mc          	& 205    	& 284         	& 0    	& 0     	\\ \hline
memory      	& 120    	& 72          	& 0    	& 0     	\\ \hline
pipeline\_logic & 406    	& 203         	& 0    	& 0     	\\ \hline
plic        	& 84     	& 1           	& 0    	& 0     	\\ \hline
prom        	& 2      	& 0           	& 0    	& 2     	\\ \hline
reg         	& 3853   	& 3398        	& 291  	& 0     	\\ \hline
CSR         	& 3181   	& 2406        	& 35   	& 0     	\\ \hline
GPR         	& 640    	& 992         	& 256  	& 0     	\\ \hline
rexmem      	& 77     	& 172         	& 0    	& 0     	\\ \hline
ridex       	& 137    	& 296         	& 0    	& 0     	\\ \hline
rifid       	& 80     	& 405         	& 0    	& 0     	\\ \hline
rmemwb      	& 87     	& 166         	& 0    	& 0     	\\ \hline
sel\_plic\_clint& 50     	& 99          	& 0    	& 0     	\\ \hline
timer       	& 245    	& 134         	& 0    	& 0     	\\ \hline
\end{tabular}
}
\caption{Resources utilization of the OBC with the PMU enabled.}
\label{tab:utilizationPMU}
\end{table}

\begin{table}[htb!]
\centering
\resizebox{.95\linewidth}{!}{
\begin{tabular}{|c|c|c|c|c|}
\hline
Component   	& Slice LUTs & Slice Registers & F7 Muxes & Block RAM \\ \hline
RV32        	& 4817   	& 4246        	& 259  	& 2     	\\ \hline
clint       	& 76     	& 72          	& 0    	& 0     	\\ \hline
decode      	& 327    	& 44          	& 0    	& 0     	\\ \hline
execute     	& 634    	& 0           	& 1    	& 0     	\\ \hline
fu          	& 74     	& 0           	& 0    	& 0     	\\ \hline
if          	& 22     	& 35          	& 0    	& 0     	\\ \hline
mc          	& 174    	& 280         	& 0    	& 0     	\\ \hline
memory      	& 118    	& 72          	& 0    	& 0     	\\ \hline
pipeline\_logic & 381    	& 203         	& 0    	& 0     	\\ \hline
plic        	& 88     	& 1           	& 0    	& 0     	\\ \hline
prom        	& 2      	& 0           	& 0    	& 2     	\\ \hline
reg         	& 2257   	& 2342        	& 258  	& 0     	\\ \hline
CSR         	& 1585   	& 1350        	& 2    	& 0     	\\ \hline
GPR         	& 640    	& 992         	& 256  	& 0     	\\ \hline
rexmem      	& 67     	& 159         	& 0    	& 0     	\\ \hline
ridex       	& 129    	& 283         	& 0    	& 0     	\\ \hline
rifid       	& 82     	& 104         	& 0    	& 0     	\\ \hline
rmemwb      	& 80     	& 152         	& 0    	& 0     	\\ \hline
sel\_plic\_clint& 50     	& 99          	& 0    	& 0     	\\ \hline
timer       	& 245    	& 134         	& 0    	& 0     	\\ \hline
\end{tabular}
}
\caption{Resources utilization of the OBC with the PMU disabled.}
\label{tab:utilization-without-PMU}
\end{table}

Table \ref{tab:utilization-relative} shows the relative difference between each design and the total resources of the FPGA used during the evaluation process \cite{digilent_nexys_2016}. Here, it shows an increase of 2.65\% in LUT logic resources and 0.85\% in Slice Registers used with the PMU enabled. Though a notable increment, it is to be expected due to the significant amount of logical resources required to control the accountability of the events. This is because, in comparison with the GPR registers, here the VHDL optimizer does not use as many F7 multiplexers resources to access the CSR registers, due to the higher complexity for accessing them, having to use LUT logic instead. On the other hand, the increment in Slice Registers is significantly smaller and justified by the higher number of registers to accommodate the PMU counters.

\begin{table}[htb]
\centering
\resizebox{\linewidth}{!}{
\begin{tabular}{|c|c|c|c|c|}
\hline
                  	& Slice LUTs & Slice Registers & F7 Muxes & Block RAM \\ \hline
Total resources board & 63400  	& 126800      	& 31700	& 135   	\\ \hline
RV32Xtrace PMU    	& 6498   	& 5347        	& 292  	& 2     	\\ \hline
\% RV32Xtrace PMU 	& 10.25\%	& 4.22\%      	& 0.92\%   & 1.48\%	\\ \hline
RV32Xtrace NO PMU 	& 4817   	& 4246        	& 259  	& 2     	\\ \hline
\% RV32Xtrace NO PMU  & 7.60\% 	& 3.35\%      	& 0.82\%   & 1.48\%	\\ \hline
\end{tabular}
}
\caption{Relative resource utilization in relation to the maximum resources available of the FPGA board.}
\label{tab:utilization-relative}
\end{table}

Next, in regard to the power usage of the implementation, two methods were used to obtain the power requirements measurements. First, the metrics provided by the simulations from the Vivado power analysis tool were extracted. These reflect the power consumption after routing and placement for the selected FPGA chip, i.e., that of the Nexys 4-DDR board \cite{digilent_nexys_2016}. The estimates obtained indicated that the OBC implementation with PMU consumed \SI{110}{\milli\watt}, whereas without PMU it employed \SI{104}{\milli\watt}. Thus, the difference in consumption with the Artix XC7A100T-1CSG324 FPGA was \SI{6}{\milli\watt} more for the implementation with PMU. Secondly, to verify that these estimates were correct, the FPGA evaluation boards were programmed with two different IPcores, one containing the RV32Xtrace OBC with the presented PMU enabled and another with the PMU disabled. This board was then connected to a configurable power supply through which the voltage and current measurements could be observed. After this, the power usage was calculated. 


\begin{table}[htb]
\centering
\resizebox{0.87\linewidth}{!}{
\begin{tabular}{|c|c|c|c|c|}
\hline
Test    	& Voltage (V) & Current (A) & Power (W) \\ \hline
Baseline	& 5       	& 0.225   	& 1.125 	\\ \hline
Programming & 5       	& 0.250   	& 1.250 	\\ \hline
Bootloader  & 5       	& 0.375   	& 1.875 	\\ \hline
Uploading   & 5       	& 0.376   	& 1.880 	\\ \hline
Dhrystone   & 5       	& 0.426   	& 2.130 	\\ \hline
Coremark	& 5       	& 0.426   	& 2.130 	\\ \hline
\end{tabular}
}
\caption{Power measurements of the OBC with the PMU enabled.}
\label{tab:power-withPMU}
\end{table}

\begin{table}[htb]
\centering
\resizebox{0.87\linewidth}{!}{
\begin{tabular}{|c|c|c|c|c|}
\hline
Test    	& Voltage (V) & Current (A) & Power (W) \\ \hline
Baseline	& 5       	& 0.225   	& 1.125 	\\ \hline
Programming & 5       	& 0.260   	& 1.300 	\\ \hline
Bootloader  & 5       	& 0.373   	& 1.865 	\\ \hline
Uploading   & 5       	& 0.375   	& 1.875 	\\ \hline
Dhrystone   & 5       	& 0.425   	& 2.125 	\\ \hline
Coremark	& 5       	& 0.425   	& 2.125 	\\ \hline
\end{tabular}
}
\caption{Power measurements of the OBC with the PMU disabled.}
\label{tab:power-withoutPMU}
\end{table}

In Tables \ref{tab:power-withPMU} and \ref{tab:power-withoutPMU}, the results of these tests can be found. For each IPcore configuration, the following set of tests were conducted. First, the baseline current was measured with the board connected to power but without any IPcore programmed into its memory. Secondly, the board was flashed with a bootloader developed in-house based on the NEORV32 BSP \cite{nolting_neorv32_2022}. Hence, the average measurements were obtained both during the programming process and afterward when the bootloader was already running in a safe state. Lastly, the bootloader was used to upload to the board the benchmarks mentioned in Section \ref{sec:benchmarks} through its serial connection. Here, the metrics of the average power employed during each of the benchmarks, were also collected, as can again be seen in the Tables \ref{tab:power-withPMU} and \ref{tab:power-withoutPMU}.

As shown in these tables, the power requirements needed by the OBC after the development of the proposed PMU have obviously increased, although by a minimal margin. One aspect worth noting is that, in order to obtain these results, the measured current and power values were averaged, as they fluctuated depending on board temperature and other confounding variables, which explains the anomaly on the \textit{Programming} row, where the power is lower on the IPcore with PMU. Nevertheless, the rest of the measurements are consistent with the difference being negligible, with the maximum increase under all conditions due to the integration of the PMU of only \SI{10}{\milli\watt}. As can be seen, these metrics overlap with the estimates from Vivado since, although the former measure the consumption of the entire board and the latter only that of the FPGA, the relative difference is comparable.

\section{Conclusions and future work}
\label{sec:conclusions}

Computing critical systems and, in particular, space-graded systems within the scope of this article, must adhere to the strictest timing requirements to ensure its correct operation and the safety of each of their components. To ease the development of this type of systems, one tool which is usually required is a statistical unit with which to extract timing and other useful information about the behavior of the system. Common performance monitoring units present the issue that it is usually very hard to match the instructions with the events that occurred during their execution on the CPU. To solve that problem, this paper displays the detailed design of a PMU that synchronizes the event counting with the instruction execution.

The design was based on an existing pipelined RISC-V processor to which the statistical counters were added. The modification adds logic to each inter-stage register to check whether the programmed events have occurred, storing this information on a new data structure, which is fed from one register to the next. This way, the event increment only occurs after the corresponding instruction has finally been written back to the GPRs and, therefore, the match between the event and the executed instruction can be stored unequivocally.

In order to validate the design, the events triggered by several programs were calculated, and then they were executed on the on-board computer. The results obtained matched perfectly with the planned outcome, demonstrating the correct operation of the presented PMU. Moreover, the events were specially selected to help reconstruct the execution model of the OBC, as discussed in more detail in this manuscript. Thus, the ability to support new events during development was tested, demonstrating the fast and easy extensibility of the PMU with minimal development time and costs. Likewise, the CoreMark and Dhrystone benchmarks were also ported to the proposed platform and compared with the results from an external reference, obtaining the results analyzed above. This characterization of the execution model and the results presented allow us to confirm the correct behavior of the HPM proposed for this article.

Furthermore, a few ideas already exist on how to enhance the current design. One area of improvement is the creation of a new way to extract these synchronized event data, for example, via an upgraded trace mechanism. Also, several mechanisms have been proposed to improve processor performance. Additionally, to refine the data on the behavior and performance of the OBC, further benchmarks and the boot software from the EPD instrument of the Solar Orbiter mission will be ported to the platform.

\section*{Declaration of competing interest}
The authors declare that they have no known competing financial interests or personal relationships that could have appeared to influence the work reported in this paper.

\section*{Data availability}
Data will be made available on request.

\section*{Funding}

This work has been supported by two predoctoral aids: ``Design and Implementation of Leon Processor Enhancements on FPGA'' of the Youth Employment Initiative (YEI) of the European Social Fund (ESF) and the Spanish Ministerio de Ciencia, Innovación y Universidades (MCIN), under the Operational Program of Youth Employment (POEJ), grant PEJ2018-004178-A; and the contract: PRE2020-094740 under the project ``Energetic Particle Detector en Solar Orbiter: fase E, calibración y explotación de datos'' reference: PID2019-104863RB-I00, funded by the MCIN (DOI: 10.13039/501100011033), the Spanish Agencia Estatal de Investigación (AEI) and by the ESF+.


 \bibliographystyle{elsarticle-num}
 \bibliography{PrimerArticulo}





\end{document}